\pdfoutput=1  

\documentclass[aps,twocolumn,nofootinbib,superscriptaddress,floatfix,preprintnumbers]{revtex4-1}

\usepackage{amsmath, float, graphicx,amssymb,multirow,pgfplots,pgfplotstable}
\usepackage{tikz,hyperref}
\usepackage{ulem}
\usepackage{comment}

\usetikzlibrary{shapes.geometric,arrows}

\begin{document}
\title{
PASSAT: Particle Accelerator helioScopes for Slim Axion-like-particle deTection
}
  \author{Walter M. Bonivento}
\email{walter.bonivento@ca.infn.it}
\affiliation{Istituto Nazionale di Fisica Nucleare, Sezione di Cagliari,
Cittadella Universitaria di Monserrato, Monserrato (CA), Italy}
  \author{Doojin Kim}
\email{doojin.kim@tamu.edu}
\affiliation{Department of Physics, University of Arizona, Tucson, AZ 85721, USA}
\affiliation{Mitchell Institute for Fundamental Physics and Astronomy, Department of Physics and Astronomy, Texas A\&M University, College Station, TX 77843, USA}
\author{Kuver Sinha}
\email{kuver.sinha@ou.edu}
\affiliation{Department of Physics and Astronomy, University of Oklahoma, Norman, OK 73019, USA}

\preprint{
\begin{minipage}[b]{1\linewidth}
\begin{flushright}
MI-TH-1934\\
\end{flushright}
\end{minipage}
}

\begin{abstract}
We propose a novel method to search for axion-like particles (ALPs) at particle accelerator experiments. ALPs produced at the target via the Primakoff effect subsequently enter a region with a magnetic field, where they are converted to photons that are then detected. Dubbed Particle Accelerator helioScopes for Slim Axion-like-particle deTection (PASSAT), our proposal uses the principle of the axion helioscope but replaces ALPs produced in the Sun with those produced in a target material. Since we rely on ALP-photon {\it conversions}, our proposal probes light (slim) ALPs that are otherwise inaccessible to laboratory-based experiments which rely on ALP decay, and complements astrophysical probes that are more model-dependent. As a first application, we reinterpret existing data from the NOMAD experiment in light of PASSAT, and constrain the parameter space for ALPs lighter than $\sim 100~{\rm eV}$ and ALP-photon coupling larger than $\sim 10^{-4}~{\rm GeV}^{-1}$. 
As benchmarks of feasible low-cost experiments improving over the NOMAD limits, we study the possibility of re-using the magnets of the CAST and the proposed BabyIAXO experiments and  placing them at the proposed BDF facility at CERN, together with some new detectors.
We find that these realizations of PASSAT  allow for a direct probe of the parameter space for ALPs lighter than $\sim 100~{\rm eV}$ and ALP-photon coupling larger than $\sim 4\times 10^{-6}~{\rm GeV}^{-1}$, which are regions that have not been probed yet by experiments with laboratory-produced ALPs. In contrast to other proposals aiming at detecting single or two-photon only events in hadronic beam dump environments, that rely heavily on Monte Carlo simulations, the background in our proposal can be directly measured {\it  in-situ}, its suppression optimized, and the irreducible background statistically subtracted. Sensitivity evaluations with other beams will be the subject of a future paper. The measurements suggested in this paper represent an additional physics case for the BDF at CERN beyond those already proposed.
\end{abstract}

\maketitle


\newcommand{\PRE}[1]{{#1}} 
\newcommand{\ul}{\underline}
\newcommand{\del}{\partial}
\newcommand{\nbox}{{\,\lower0.9pt\vbox{\hrule \hbox{\vrule height 0.2 cm
\hskip 0.2 cm \vrule height 0.2 cm}\hrule}\,}}

\newcommand{\postscript}[2]{\setlength{\epsfxsize}{#2\hsize}
   \centerline{\epsfbox{#1}}}
\newcommand{\gweak}{g_{\text{weak}}}
\newcommand{\mweak}{m_{\text{weak}}}
\newcommand{\mplanck}{M_{\text{Pl}}}
\newcommand{\mstar}{M_{*}}
\newcommand{\sigmaan}{\sigma_{\text{an}}}
\newcommand{\sigmatot}{\sigma_{\text{tot}}}
\newcommand{\sigmaSI}{\sigma_{\rm SI}}
\newcommand{\sigmaSD}{\sigma_{\rm SD}}
\newcommand{\OmegaM}{\Omega_{\text{M}}}
\newcommand{\OmegaDM}{\Omega_{\text{DM}}}
\newcommand{\ipb}{\text{pb}^{-1}}
\newcommand{\ifb}{\text{fb}^{-1}}
\newcommand{\iab}{\text{ab}^{-1}}
\newcommand{\ev}{\text{eV}}
\newcommand{\kev}{\text{keV}}
\newcommand{\mev}{\text{MeV}}
\newcommand{\gev}{\text{GeV}}
\newcommand{\tev}{\text{TeV}}
\newcommand{\pb}{\text{pb}}
\newcommand{\mb}{\text{mb}}
\newcommand{\cm}{\text{cm}}
\newcommand{\m}{\text{m}}
\newcommand{\km}{\text{km}}
\newcommand{\kg}{\text{kg}}
\newcommand{\g}{\text{g}}
\newcommand{\s}{\text{s}}
\newcommand{\yr}{\text{yr}}
\newcommand{\Mpc}{\text{Mpc}}
\newcommand{\etal}{{\em et al.}}
\newcommand{\eg}{{\em e.g.}}
\newcommand{\ie}{{\em i.e.}}
\newcommand{\ibid}{{\em ibid.}}
\newcommand{\Eqref}[1]{Equation~(\ref{#1})}
\newcommand{\secref}[1]{Sec.~\ref{sec:#1}}
\newcommand{\secsref}[2]{Secs.~\ref{sec:#1} and \ref{sec:#2}}
\newcommand{\Secref}[1]{Section~\ref{sec:#1}}
\newcommand{\appref}[1]{App.~\ref{sec:#1}}
\newcommand{\figref}[1]{Fig.~\ref{fig:#1}}
\newcommand{\figsref}[2]{Figs.~\ref{fig:#1} and \ref{fig:#2}}
\newcommand{\Figref}[1]{Figure~\ref{fig:#1}}
\newcommand{\tableref}[1]{Table~\ref{table:#1}}
\newcommand{\tablesref}[2]{Tables~\ref{table:#1} and \ref{table:#2}}
\newcommand{\Dsle}[1]{\slash\hskip -0.28 cm #1}
\newcommand{\met}{{\Dsle E_T}}
\newcommand{\mpt}{\not{\! p_T}}
\newcommand{\Dslp}[1]{\slash\hskip -0.23 cm #1}
\newcommand{\Dsl}[1]{\slash\hskip -0.20 cm #1}

\newcommand{\mB}{m_{B^1}}
\newcommand{\mq}{m_{q^1}}
\newcommand{\mf}{m_{f^1}}
\newcommand{\mKK}{m_{KK}}
\newcommand{\WIMP}{\text{WIMP}}
\newcommand{\SWIMP}{\text{SWIMP}}
\newcommand{\NLSP}{\text{NLSP}}
\newcommand{\LSP}{\text{LSP}}
\newcommand{\mWIMP}{m_{\WIMP}}
\newcommand{\mSWIMP}{m_{\SWIMP}}
\newcommand{\mNLSP}{m_{\NLSP}}
\newcommand{\mchi}{m_{\chi}}
\newcommand{\mgravitino}{m_{\gravitino}}
\newcommand{\mmed}{M_{\text{med}}}
\newcommand{\gravitino}{\tilde{G}}
\newcommand{\Bino}{\tilde{B}}
\newcommand{\photino}{\tilde{\gamma}}
\newcommand{\stau}{\tilde{\tau}}
\newcommand{\slepton}{\tilde{l}}
\newcommand{\snu}{\tilde{\nu}}
\newcommand{\squark}{\tilde{q}}
\newcommand{\mgaugino}{M_{1/2}}
\newcommand{\epsEM}{\varepsilon_{\text{EM}}}
\newcommand{\mmess}{M_{\text{mess}}}
\newcommand{\lmess}{\Lambda}
\newcommand{\nmess}{N_{\text{m}}}
\newcommand{\signmu}{\text{sign}(\mu)}
\newcommand{\Omegachi}{\Omega_{\chi}}
\newcommand{\lambdafs}{\lambda_{\text{FS}}}
\newcommand{\be}{\begin{equation}}
\newcommand{\ee}{\end{equation}}
\newcommand{\bea}{\begin{eqnarray}}
\newcommand{\eea}{\end{eqnarray}}
\newcommand{\beq}{\begin{equation}}
\newcommand{\eeq}{\end{equation}}
\newcommand{\beqn}{\begin{eqnarray}}
\newcommand{\eeqn}{\end{eqnarray}}
\newcommand{\baln}{\begin{align}}
\newcommand{\ealn}{\end{align}}
\newcommand{\lsim}{\lower.7ex\hbox{$\;\stackrel{\textstyle<}{\sim}\;$}}
\newcommand{\gsim}{\lower.7ex\hbox{$\;\stackrel{\textstyle>}{\sim}\;$}}

\newcommand{\ssection}[1]{{\em #1.\ }}
\newcommand{\rem}[1]{\textbf{#1}}

\def\ie{{\it i.e.}\/}
\def\eg{{\it e.g.}\/}
\def\etc{{\it etc}.\/}
\def\calN{{\cal N}}

\def\mptwo{{m_{\pi^0}^2}}
\def\mp{{m_{\pi^0}}}
\def\sqtsn{\sqrt{s_n}}
\def\sqtsn{\sqrt{s_n}}
\def\sqtsn{\sqrt{s_n}}
\def\sqts0{\sqrt{s_0}}
\def\Dsqts{\Delta(\sqrt{s})}
\def\Omegatot{\Omega_{\mathrm{tot}}}

\newcommand{\changed}[2]{{\protect\color{red}\sout{#1}}{\protect\color{blue}\uwave{#2}}}


\section{Introduction and Motivations}
The QCD axion \cite{Weinberg:1977ma, Wilczek:1977pj, Peccei:1977hh} and more general pseudo-scalar axion-like-particles (ALPs), which are ubiquitous in string theory \cite{Arvanitaki:2009fg, Cicoli:2012sz}, are a major focus of searches for physics beyond the Standard Model (SM). The experimental ecosystem investigating these particles is vast and rich. The techniques often rely on the ALP-photon coupling:
\beq
\mathcal{L}_{\rm int} \supset -\frac{1}{4} g_{a\gamma\gamma}a F^{\mu\nu}\tilde{F}_{\mu\nu}\,,
\eeq
where $g_{a\gamma\gamma}$ is the coupling between ALP (henceforth denoted by $a$) and the SM photon and where $F_{\mu\nu}$ ($\tilde{F}_{\mu\nu}$) is the usual field (dual field) strength tensor of the photon. ALPs can convert to photons and vice versa in the presence of an external magnetic field, leading to experiments based on the axion haloscope and the axion helioscope~\cite{Sikivie:1983ip,Sikivie:1985yu,Anastassopoulos:2017ftl}. On the other hand, photon regeneration (light shining through wall, LSW) experiments~\cite{Spector:2019ooq} attempt to actively produce ALPs with a high-intensity laser beam applied in a magnetic field, followed by detecting the 
produced photons via ALP-photon conversion.  We refer to Ref.~\cite{Graham:2015ouw} for a recent review of these topics.

In contrast, particle accelerator-based experiments have so far been proposed to search for heavy (i.e., MeV -- GeV-scale) ALPs, since they are relatively short-lived and their decay point is not much displaced from their production point.   
Examples include NA62~\cite{Brunetti:2018ffq}, SHiP~\cite{Alekhin:2015byh}, FASER~\cite{Ariga:2018pin,Ariga:2019ufm}, and SeaQuest~\cite{Berlin:2018pwi}. In such experiments, an ALP is first produced in the target/dump material and subsequently decays to  photons in the decay volume.  As the ALP mass is lowered, it exits the decay volume without decaying and these experiments lose their sensitivity. We refer to \cite{Dobrich:2015jyk, Dobrich:2019dxc, Bauer:2018uxu, Feng:2018noy} for recent theoretical studies on ALP searches. 

\begin{figure}[t]
\centering
\includegraphics[width=8.4cm]{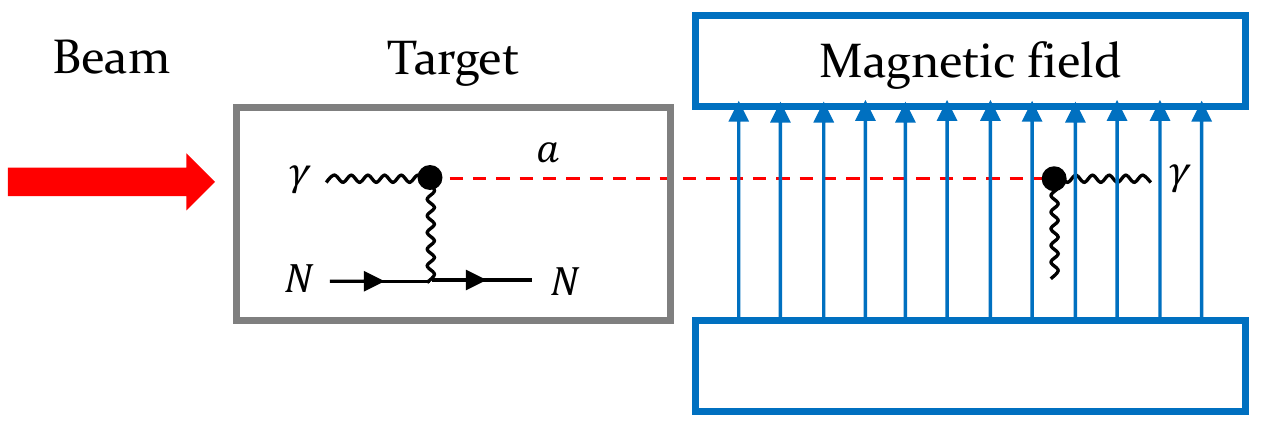}
\caption{\label{fig:setup} Schematic description of the PASSAT search strategy under consideration}
\end{figure}

The purpose of this paper is to point out that if the ALP enters a region with a transverse magnetic  field after being produced at the target, then a beam dump becomes sensitive to very light ALPs (see Figure~\ref{fig:setup}). This is because the ALP is no longer required to decay; rather, the ALP \textit{converts} to a photon which can be detected. If the length traversed by the ALP is shorter than the associated oscillation length, the conversion is coherent and a net probability of conversion can be obtained as a function of the ALP-photon coupling. The predicted photon signal is a product of the ALP production cross section at the target material and its conversion probability as it subsequently traverses the magnetic field.

Dubbed Particle Accelerator helioScopes for Slim Axion-like-particle deTection (PASSAT), our proposal thus combines the principle of the axion helioscope with traditional ALP production in beam dumps. From the perspective of a traditional helioscope, the Sun is replaced by the target material as the source of ALPs; from the perspective of a traditional beam dump experiment, the ALP decay process is replaced by the ALP conversion process. 

While its expected reach cannot rival that of the CAST helioscope, PASSAT outperforms other searches for  \textit{laboratory-produced} ALPs. 
We stress that it is important to pursue laboratory-based experiments to probe the ALP parameter space, even in regimes that are constrained by helioscopes and astrophysical sources. As pointed out in the aftermath of the PVLAS signal~\cite{Zavattini:2005tm}, there are many assumptions that lie behind constraints given by solar helioscopes (we choose \cite{Jaeckel:2006xm, Ahlers:2006iz, Brax:2007ak} among the many papers that make this point). The environmental conditions for the production of ALPs inside the Sun or in stars are very different from those in laboratories. The coupling $g_{a \gamma \gamma}$ or the mass $m_a$ can depend on a host of environmental parameters, such as  the temperature, matter density, or plasma frequency, as well as the momentum transfer at the ALP-photon vertex.
Thus, laboratory-based searches, apart from being complementary to astrophysical searches, are also more \textit{conservative}.

We begin our work by discussing ALP production and conversion at PASSAT, and then apply the resulting formalism to several implementations in past, current, and future experimental facilities to obtain an initial estimate of the expected performance. We first reinterpret existing data from the NOMAD experiment~\cite{Astier:2000gx} in light of PASSAT. We then discuss the possible future implementation of PASSAT at CERN, combining two components: $(i)$ a permanent fixed target complex composed of the beam line and target infrastructure from the proposed Beam Dump Facility (BDF)~\cite{Kershaw:2018pyb} and the muon shield component from the SHiP detector~\cite{Anelli:2015pba} $(ii)$ followed by the CAST~\cite{Zioutas:1998cc} or BabyIAXO~\cite{Armengaud:2019uso} magnets, together with some new detectors. 
In each case, we provide the experimental parameters and the projected sensitivities, including  the future potential of upgrading LHC-like magnets up to 20~T, as projected by studies related to the FCC-hh collider. 

\section{ALP Production and Conversion} 
In this section, we calculate the expected number of events $N_{\rm ex}$ from ALP production via the Primakoff process at the target followed by its conversion to a photon in the magnetic field.
The quantity $N_{\rm ex}$ is simply given as 
\beq
N_{\rm ex} =  N_{\rm POT}\cdot \frac{1}{\sigma_{\gamma \to {\rm all}}}\int dE_a d\theta_a \frac{d^2\sigma_a}{dE_ad\theta_a} \cdot P_{a\to\gamma} \cdot P_{\rm surv} \,, \label{eq:Nex}
\eeq
where $N_{\rm POT}$ is the total number of protons on target (POT) and $\sigma_{\gamma \to {\rm all}}$ is the cross section for photon-nucleus scattering which is dominated by photon conversion to an electron-positron pair in the nuclear fields.
For the photon energies of our interest, we have $\sigma_{\gamma \to {\rm all}} \approx 1.2\times 10^4$ mb in molybdenum and $\sigma_{\gamma \to {\rm all}} \approx 1.4\times 10^2$ mb in beryllium~\cite{Tanabashi:2018oca}.

A couple of probabilities are involved in this signal rate calculation. First, $P_{a\to\gamma}$ stands for the probability of ALP-to-photon conversion when an ALP travels distance $L$ in a magnetic field $B$:
\beq
P_{a\to\gamma}=\left(\frac{g_{a\gamma\gamma}BL}{2} \right)^2 \left(\frac{2}{qL}\right)^2 \sin^2 \left(\frac{qL}{2} \right)\,, \label{eq:pconv}
\eeq
where the product of the second and third factors is the form factor reflecting the coherence of the conversion. In the relativistic limit and in vacuum, $q$ is given by 
\begin{equation}
q \, = \,  2\sqrt{\left(\frac{m^2_a}{4 E_a} \right)^2 + \left(\frac{1}{2} g_{a \gamma \gamma} B \right)^2 }.
\end{equation} 
The other one $P_{\rm surv}$ describes the survival probability that an ALP reaches the detector before decaying into a photon pair. The usual decay law suggests
\begin{equation}
    P_{\rm surv}=\exp\left(-\frac{\tilde{L}}{L_a^{\rm lab}} \right)\,,
\end{equation}
where $\tilde{L}$ is the distance between the target and the detector and $L_a^{\rm lab}$ is the laboratory-frame mean decay length of the ALP. In terms of the Lorentz boost factor of ALP $\gamma_a$ and the decay width of ALP $\Gamma_a$, $L_a^{\rm lab}$ is given by
\begin{equation}
    L_a^{\rm lab}= \frac{\sqrt{\gamma_a^2-1}\,c}{\Gamma_a}\,,
\end{equation}
where $c$ is the speed of light and $\Gamma_a$ associated with the diphoton mode is
\begin{equation}
    \Gamma_a =\frac{g_{a\gamma\gamma}^2 m_a^3}{64\pi}\,.
\end{equation}

The integrand in Eq.~\eqref{eq:Nex} describes the differential ALP production cross section via the Primakoff process convoluted with a differential photon number density profile $n_\gamma$, in ALP energy $E_a$ and its outgoing angle $\theta_a$ from the beam axis. This is given by
\beq
\frac{d^2\sigma_a}{dE_ad\theta_a} = \int dp_T^2d\phi~ n_\gamma(E_a,p_T^2)\frac{d\sigma_{\gamma N}}{d\theta_a}\,, \label{eq:dsdedt}
\eeq
where $\phi$ denotes the angle in the transverse plane between the incoming photon and the outgoing ALP  and  $\sigma_{\gamma N}$ is the cross section for the Primakoff process. We note that $n_\gamma$ is a function over the photon transverse momentum $p_T$ and we have substituted the $E_\gamma$ dependence in $n_\gamma$ with $E_a$ in the collinear limit, i.e., $E_a \approx E_\gamma$.

In the massless (or ultra relativistic) ALP limit, the $\sigma_{\gamma N}$ is approximately of the form~\cite{Dobrich:2015jyk,Feng:2018noy}
\beq
\frac{d\sigma_{\gamma N}}{d\theta_a}\approx -\frac{1}{16}\alpha g_{a\gamma\gamma}^2 Z^2 F(|t|)^2 \frac{(4E_a^2t+m_a^4)}{t^2}\theta_a\,,
\eeq 
where $\alpha$ and $Z$ are the fine structure constant and the atomic number of target material, respectively, and where $t=-(p_\gamma-p_a)^2$ is given by
\beq
t = -\frac{m_a^4}{4 E_a^2}-p_T^2+2E_a \sqrt{p_T^2}\theta_a \cos\phi -E_a^2\theta_a^2\,.
\eeq
We follow Ref.~\cite{Dobrich:2015jyk} and choose the Helm form factor as $F(|t|)$ which is assumed to be vanishing for $\sqrt{|t|}R_1>4.49$:
\beq
F(|t|)=\frac{3j_1(\sqrt{|t|}R_1)}{\sqrt{|t|}R_1}\exp\left( -\frac{|t|s^2}{2}\right)\Theta(4.49-\sqrt{|t|}R_1)\,,
\eeq
where $s=0.9$~fm, $\Theta(x)$ is the Heaviside step function, and $j_1$ is the spherical Bessel function of the first kind. 
Here $R_1$ is parametrized according to Ref.~\cite{Lewin:1995rx}, i.e., $R_1=\sqrt{(1.23 A^{1/3}-0.6)^2+2.18}$~fm with $A$ being the atomic mass number of target material.

We finally turn to the photon number density profile $n_\gamma$. In principle, the precise determination of $n_\gamma$ requires a full detector-level  simulation. However, in this first study, we opt to take a (semi-)analytic approach, based on an empirical model. This enables us to perform rapid estimates, essentially as a proof of principle.  

Since the ALPs under consideration are sufficiently light, the dominant photon source is the decay of mesons (e.g., $\pi^0, \eta$). 
Production of mesons through high-energy proton beams on target is elegantly parametrized by the so-called BMPT model~\cite{Bonesini:2001iz} whose fits were tuned with $p$-beryllium target collision data.
In the limit of negligible transverse momenta of mesons, we find that the differential production cross section of say, a charged pion has the form of
\begin{eqnarray}
\frac{d\sigma}{dE_\pi}&\propto& E_\pi \left(1-\frac{E_\pi}{E_{\rm beam}}\right)^{c_\alpha}\left(1+c_\beta\frac{E_\pi}{E_{\rm beam}}\right)\nonumber \\
&\times& \left(\frac{E_\pi}{E_{\rm beam}}\right)^{-c_\gamma}\,, 
\end{eqnarray}
where $E_{\rm beam}$ denotes the particle beam energy.
We assume that the functional behavior of the above spectrum describes the differential $\pi^0$ number density $dN_\pi/dE_\pi$. Our parameter choices $(c_\alpha,c_\beta,c_\gamma)$ for $\pi^\pm$ are $(3.45,1.57,0.517)$~\cite{Bonesini:2001iz}. 
We further assume that $N_\pi$ is normalized to 4 for $E_{\rm beam}=400$ GeV considering the measurement data in Ref.~\cite{Aguilar-Benitez:216837} and the simulation study in Ref.~\cite{Dobrich:2019dxc}.
The pion energy $E_\pi$ and the photon energy $E_\gamma$ in the laboratory frame are related by the following Lorentz transformation, 
\beq
E_\gamma = E_\gamma^*(\gamma_\pi +\sqrt{\gamma_\pi^2-1}\cos\theta_\gamma^*)\,,
\eeq
where $E_\gamma^*(=m_\pi/2)$ and $\theta_\gamma^*$ are the photon energy and the photon emission angle in the pion rest frame, respectively, and where the pion boost factor contains $E_\pi$ dependence such that $\gamma_\pi = E_\pi /m_\pi$. 
Since typical pions are highly boosted, $\sqrt{\gamma_\pi^2-1}\approx \gamma_\pi$ so that we have
\beq
E_\pi \approx \frac{m_\pi}{E_\gamma^*+\sqrt{E_\gamma^{*2}-p_T^2}}E_\gamma\,. 
 \label{eq:Erel}
\eeq
The density of $p_T^2$, $w(p_T^2)$, can be calculated from the density of $\cos\theta_\gamma^*$, $w(\cos\theta_\gamma^*)$, that is, 
\beq
w(p_T^2)=\left|\frac{d\cos\theta_\gamma^*}{dp_T^2}\right| w(\cos\theta_\gamma^*)=\frac{1}{4E_\gamma^*\sqrt{E_\gamma^{*2}-p_T^2}}\,.
\eeq
Here $w(\cos\theta_\gamma^*)=1/2$ as only the forward-moving photon contributes to the final estimate out of the two decay products. 
Taking the collinear limit $E_a \approx E_\gamma$ again, therefore, we find
\beq
n_\gamma(E_a,p_T^2)=\left|\frac{dE_\pi}{dE_\gamma} \right|\frac{dN_\pi}{dE_\pi}w(p_T^2)\,,
\eeq
where the Jacobian factor can be readily computed from Eq.~\eqref{eq:Erel}.

\section{Reinterpretation of past experiments}
We first discuss the NOMAD experiment since it has experimental data and possesses the main features of PASSAT, a fact that was not appreciated in the past. 
\begin{table}[t]
\centering
\begin{tabular}{c| c c c c}
\hline \hline 
Exp. & $B$ [Tesla] & $L$ [m] & $A$ [cm$^2$] & $\theta_a^{\max}$ [mrad]  \\
\hline
NOMAD~\cite{Vannucci:1666456} & 0.4 & 7.5 & $3.8\times 10^4$ & 2.1 \\
\hline
CAST~\cite{Zioutas:1998cc} & 8.4 & 9.26 & 14.5 & 0.36\\
BabyIAXO~\cite{Armengaud:2019uso} & 2 & 10 & $7.7\times 10^3$ & 8.3\\
\hline \hline
\end{tabular}
\caption{\label{tab:spec} Experimental parameters for the magnetic field area of the benchmark experiments. 
The strength of $B$ for BabyIAXO is the claimed average value. 
The maximum angular acceptance $\theta_a^{\max}$ is calculated with respect to the full distance between the target and the photon detector. See the text for details. }
\end{table}
The ALP search at NOMAD~\cite{Astier:2000gx} assumes that a fraction of the  photons produced by a proton beam on the beryllium target enters the horn region where they are converted to ALPs. These ALPs are subsequently re-converted back to photons in the NOMAD spectrometer. The underlying principle is the LSW class of experiments, with the laser replaced by a beam dump. Like a LSW experiment, it requires two conversion stages in two separate regions with magnetic fields: first from photons to ALPs, and then ALPs back to photons.

Motivated by the idea of PASSAT applied to NOMAD, we instead focus on the ALPs directly created in the target, jettisoning the first phase of photon to ALP conversion. ALPs directly produced in the target by the Primakoff process will subsequently be converted back to photons in the magnetic field region of the NOMAD spectrometer. The distance between the target and the detector is 835 m, and in the conversion region a 0.4~T magnetic field is applied for a length $L=7.5$~m. The width$\times$height of the magnetic field region is $3.5\times3.5$~m$^2$. 

We take a circular area of radius 1.75~m instead of the square cross section for convenience of calculation. The produced ALPs should reach the photon detector without being absorbed in other parts of experimental facility. This imposes a maximum allowed value of $\theta_a$, given  approximately by the ratio of radius of cross-sectional area $A$ to the entire distance between the target and the ($\gamma$-ray) detector (see also Table~\ref{tab:spec}). Thus, the maximum angular acceptance for ALPs in NOMAD-PASSAT is given simply by $\theta_a^{\max}$ $\approx$ $1.75/(835+7.5)\approx 2.1$ mrad. The values of other key parameters for NOMAD-PASSAT are summarized in Table~\ref{tab:spec}.

\section{Possible new experiments}
As far as future experiments are concerned, we propose to recycle the  magnets from CAST or BabyIAXO experiments, after they are decommissioned, and locate them at the BDF complex, possibly after its first use with the SHiP experiment. The BDF project, which is currently in its planning phase, will be housed  in the North Area of CERN’s Prevessin site and utilize the 400~GeV Super Proton Synchrotron (SPS) proton beam. This component will provide the beam line and target infrastructure. It is to this component that the muon shield from SHiP will be added, forming the BDF-SHiP complex.

\begin{figure}[t]
\centering
\includegraphics[width=8.4cm]{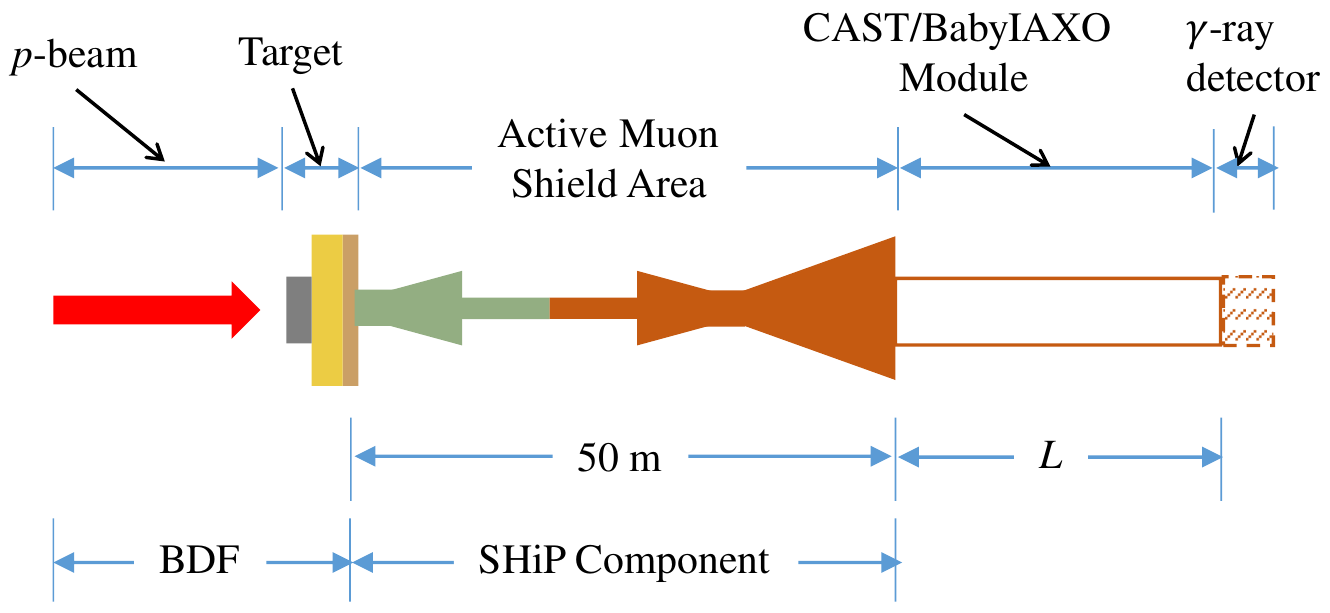}
\caption{\label{fig:prop} Conceptual design of PASSAT with the CAST or BabyIAXO magnets. 
The components from the proton beam through the active muon shield area are from the BDF-SHiP complex proposed at CERN. The X-ray-sensitive detector of CAST/BabyIAXO is replaced by a $\gamma$-ray detector. Some additional veto detectors for background suppression may be needed as well.
 }
\end{figure}

The conceptual design of the experimental setup that we are envisioning is depicted in Figure~\ref{fig:prop}. The BDF/SHiP module includes the proton beam, target complex, and active muon shield. In our study we assume that the beam energy is 400~GeV and the target in the core of the shower is molybdenum (which is contained in TZM alloy) like BDF for purposes of illustration. The produced ALPs traverse the muon shield area and enter the bore of CAST/BabyIAXO where the ALP-photon conversion occurs.

We note that the iron-filled active muon shield area itself comes with a 50~m-long transverse magnetic field region of approximately 1.8~T. Some ALPs may convert to photons in the magnetic field in the shield area, but such photons will be quickly absorbed to the material. We thus neglect any contribution to the final photon count arising from conversions in the muon shield area. A fraction $\sim 10^{-7}$ of incident ALPs are lost due to conversion in the muon shield in this manner.

We make a few comments on the input parameters required to calculate the final expected number of photons. Firstly, POT is expected to be $2 \times 10^{20}$ in five years. Secondly, when calculating the maximum angular acceptances, the length of the muon shield area should be taken into account; for example, $\theta_a^{\max}$ for the BabyIAXO case is $0.5/(50+10)\approx 8.3$ mrad. The values of the magnetic field and lengths of the different conversion modules are tabulated in Table~\ref{tab:spec}.  Finally, we note that the photon detectors in the CAST/BabyIAXO experiments are designed to be sensitive to X-ray. 
For our purpose, the detector is replaced by a $\gamma$-ray photon detector (e.g., calorimeter). 

\section{Experimental Sensitivities}

From Eq.~\eqref{eq:Nex}, the expected number of events can be obtained by integrating the differential cross section in Eq.~\eqref{eq:dsdedt} over $E_a$, $\theta_a$, $\phi$ and $p_T^2$. Firstly, in all our sensitivity calculations, we impose the requirement that $E_a>50$ GeV. 
Given the fact that the produced mesons are forward-directed and not much transverse~\cite{Bonesini:2001iz}, this implies that the momenta of incoming photons and in turn their parent mesons along the beam axis dominate so significantly that one can neglect their transverse momentum. 
In other words, we restrict our initial estimates to the phase space where the negligible transverse momentum approximation and the limit of ultra relativistic mesons are sufficiently valid. 
The unconsidered phase space can provide an additional contribution to the signal sensitivities of interest, and therefore, our estimate here may be understood as being rather conservative. 
We will perform a dedicated study including full experimental setup details and the (significant) cascade factor of showering in future work~\cite{future}.

The photons from  ALP conversion in the magnetic field should reach the photon detector without being absorbed in other parts of experimental facility, giving a maximum angular acceptance $\theta_a$ for each benchmark experiment, summarized in  Table~\ref{tab:spec}. 
$\phi$ simply ranges from 0 to $\pi$. 
Finally, $p_T^2$ spans 0 to $m_\pi^2/4$. 
Since we are restricted to $E_a>50$ GeV, the full $p_T$ range is within their angular acceptance. 
However, the angular acceptance for CAST is somewhat limited, so we consider $p_T^2 \in (0, 36^2)~{\rm MeV}^2$ in the corresponding calculation.

\begin{figure}[t]
\centering
\includegraphics[width=8.4cm]{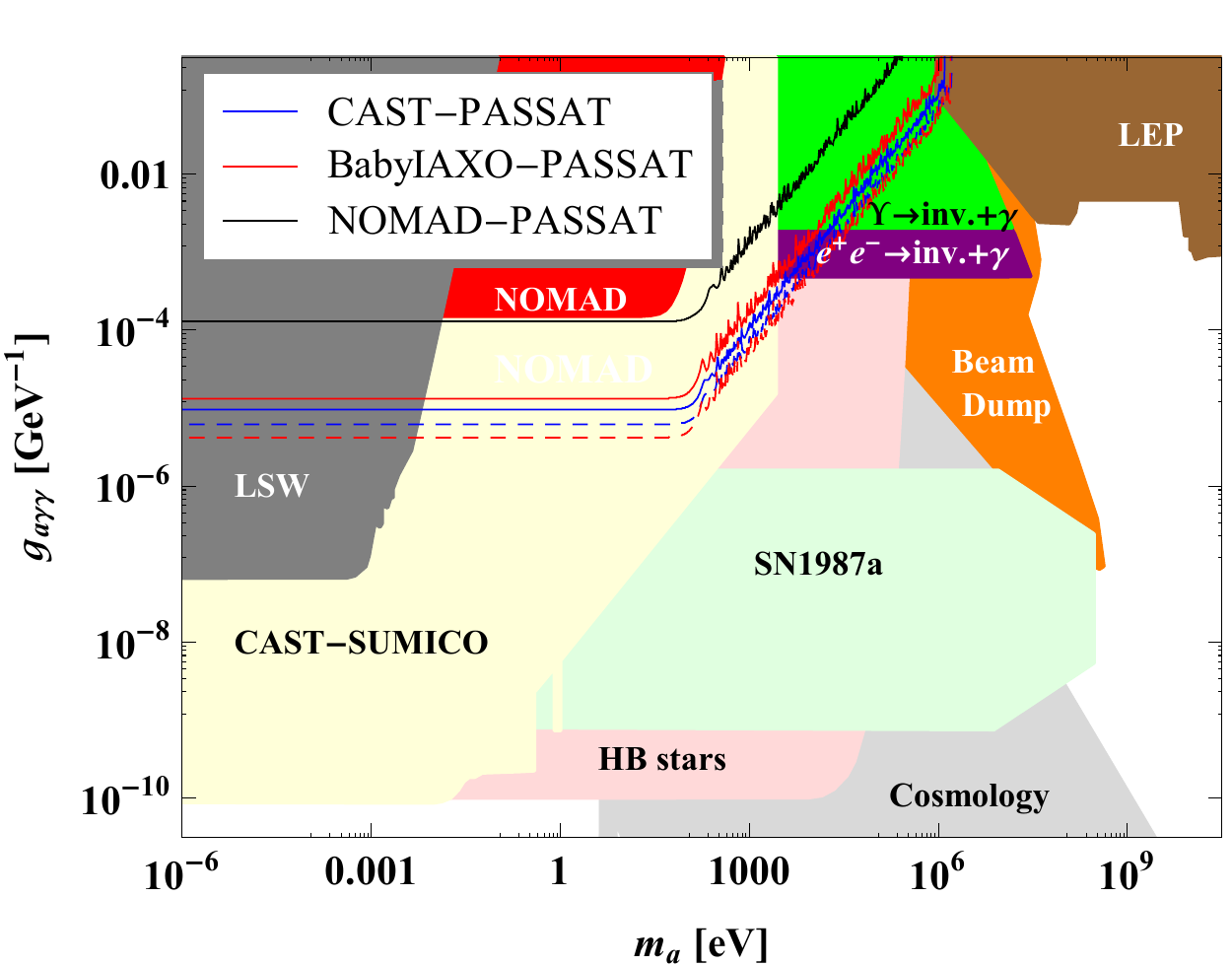}
\caption{\label{fig:expsense} Expected experimental sensitivity that can be achieved by PASSAT with CAST (blue lines) and BabyIAXO (red lines) magnets, in the plane of ALP mass $m_a$ and the associated photon coupling $g_{a\gamma\gamma}$.
The NOMAD ALP search in~\cite{Astier:2000gx} is reinterpreted in terms of PASSAT as the experiment possesses the main features of PASSAT, and the black solid line shows the resulting limit. 
Dashed lines are the corresponding sensitivities with prospective $B$ of 20~T for CAST-PASSAT and BabyIAXO-PASSAT.  
The current bounds summarized in e.g.,  Refs.~\cite{Jaeckel:2015jla,Bauer:2018uxu} are from existing ALP searches. 
The prospective limits are estimated by 90\% C.L. under the assumptions of 10 (negligible) background events for BabyIAXO-PASSAT (CAST-PASSAT) and $2\times 10^{20}$ POT. 
The NOMAD case is calculated with the expected number of background events reported in~\cite{Astier:2000gx}, being normalized to $1.08\times 10^{19}$ POT.
}
\end{figure}

The curves in Figure~\ref{fig:expsense} display our estimate of the experimental sensitivity and existing bounds from various laboratory-produced ALP search experiments (LSW, NOMAD, $\Upsilon\to {\rm inv.}+\gamma$, $e^+e^- \to {\rm inv.}+\gamma$, Beam Dump, and LEP) and astrophysical/cosmological searches (CAST-SUMICO, HB stars, SN1987a, and Cosmology) in the plane of $m_a$ and $g_{a\gamma\gamma}$. 
We first estimate the exclusion limit that the NOMAD experiment would reach with existing data. 
With a 450 GeV proton beam on the beryllium target and $1.08 \times 10^{19}$ POT, no significant excess has been observed over the expected neutrino background $272 \pm 18$ events~\cite{Astier:2000gx} which can occur in the preshower region or in the upstream region.
We consider $E_a \approx E_\gamma$ ranging 50 GeV to 140 GeV, while conservatively assuming that all expected background events are relevant to this energy range. 
The limit is computed at 90\% C.L. for a given background assumption and its statistical uncertainty.\footnote{Since the beryllium target in NOMAD is as thin as 100 mm, some fraction of incident protons may not scatter in the target but traverse towards the downstream dump area. We here take a rather simple analysis scheme, not distinguishing ALPs produced at the target with those produced at the downstream complex, which is beyond the scope of this study.}
The result is shown by the black-solid curve in Figure~\ref{fig:expsense}. The red-shaded region denotes the bounds published by the NOMAD Collaboration, assuming a LSW interpretation. We see that NOMAD-PASSAT constrains a wider range of parameter space, not only covering the existing NOMAD bound but exploring up to $g_{a\gamma\gamma}\sim 10^{-4}~{\rm GeV}^{-1}$ for $m_a \lesssim 0.1$ keV. 

When it comes to the other possible experiments, we assume that 400 GeV of CERN SPS proton beam is incident on a molybdenum target with specifications similar to those of the target adopted in the BDF project, as mentioned earlier.
We include the contribution from the $\eta$ meson decay, simply assuming that the expected number of photons is roughly 1/10 of the $\pi^0$ meson case~\cite{Dobrich:2019dxc,Aguilar-Benitez:216837}, although the experimental sensitivities are not much affected by this inclusion. 
Again, the limits are calculated by 90\% C.L. with $2\times 10^{20}$ POT that BDF aims to achieve~\cite{Kershaw:2018pyb}.

\section{Considerations on background}

At the expected proton beam intensities at the BDF, there will still be a large flux of muons coming out of the active muon shield. Therefore, to suppress background from charged particles and possible neutral particles at the exit point of the shield, a set of properly optimized veto detectors will be needed. To avoid interactions in the bore region of the CAST/BabyIAXO module some level of vacuum could also be needed. 

There will be an irreducible  background arising from $\nu_e$ elastic scattering (ES) events, mostly coming from the decays of charmed particles in the beam dump, inside the photon detector. We attempt an order of magnitude estimate here based on previous related studies that can be found in~\cite{Anelli:2015pba} and subsequent documents for the 9,600~kg neutrino detector of the SHiP experiment. The number of ES found there can be scaled to what is expected in PASSAT from the detector mass ratio.
We assume for PASSAT a calorimeter made of $(i)$ an electromagnetic section with appropriate longitudinal segmentation for selecting interactions in the first four radiation lengths, in order to keep high efficiency on signal photons and good angular resolution and $(ii)$ a hadronic section aiming at rejecting deep inelastic scattering events. A possible implementation of such a calorimeter was first described in~\cite{Bonivento:2018eqn} and is now the baseline option for the proposed SHiP experiment.

Assuming that the neutrino flux is uniformly incident on the calorimeter, the expected number of ES events at BabyIAXO-PASSAT would be 17, obtained as follows: 800 (the number of background events for a dark matter search with electron scattering in~\cite{Anelli:2015pba})~$\times~200$ (weight of a 4X$_0$ lead calorimeter of BabyIAXO magnet area)~$/~9,600 \approx 17$. 
The requirement of $E_a\approx E_\gamma>50$~GeV further reduces $80\%-90$\% of the ES background~\cite{Anelli:2015pba}. 
In this analysis we conservatively assume 10 background events for BabyIAXO-PASSAT. On the other hand, for CAST-PASSAT we assume negligible background as the associated cross sectional area $A$ is much smaller, implying an even more suppressed neutrino flux entering the calorimeter.  

It should be stressed that, in contrast to other proposed experiments at  beam dumps looking for New Physics particles decaying to neutral particles only, such as~\cite{Magill:2018jla} and~\cite{Harland-Lang:2019zur}, for PASSAT it is possible to determine the background directly from the data, by running with the magnet current switched off. It is reasonable that the data acquisition can include periods with current on and off for equal duration. This will also allow us to optimize the setup and evaluate its feasibility at the beginning of the experiment.

\section{Results}

Our analysis results with CAST and BabyIAXO magnets are exhibited by the blue and red solid curves, respectively, in Figure~\ref{fig:expsense}.
We clearly see that all of the benchmark experiments promoted to PASSAT show equally good capabilities and allow for probing a substantially broader range of parameter space than explored by past laboratory-produced ALP searches. 

Beyond $m_a\approx 0.01 - 0.1$ keV, the associated oscillation length of the produced ALP becomes shorter than $L$ and the conversion mechanism becomes non-coherent, resulting in a rising sensitivity line.
The conversion process essentially competes with the decay process. 
The probability that the produced ALPs decay before reaching the detector, $P_{\rm decay}$ is given by
\begin{equation}
    P_{\rm decay}=1-P_{\rm surv}\,.
\end{equation}
Once this probability becomes comparable to the conversion probability in Eq.~\eqref{eq:pconv}, PASSAT starts to lose the sensitivity and the signal detection via the ALP decay would appear competitive.\footnote{It is interesting {\it per se} to probe the ALP parameter space through the decay channel under the PASSAT setup. We reserve this for future work.} 
For our benchmark experiments, our numerical scan shows that it takes place around the beam dump limits.
We further show the expected experimental sensitivities with a prospective higher magnetic field of 20 T,\footnote{Although high-$B$ magnets are better incorporated to CAST-like detectors with a small aperture, we perform optimistic estimates, assuming that the same magnet technology will be available for BabyIAXO in the near future.} by the dashed but same color-coded lines. 
BabyIAXO (CAST) can accomplish sizable (mild) improvements in the sensitivities as $B$ is increased by an order of magnitude (a factor of $\sim 2$). 

\medskip

\section{Conclusions and discussion} 
We have proposed a novel method to search for ALPs at particle accelerator experiments. Our results suggest that PASSAT should probe a wide range of parameter space that none of the laboratory-produced ALP search experiments have ever explored. In particular, the expected experimental sensitivity covers regions explored by the CAST helioscope experiment, providing a conservative and complementary probe. The experimental sensitivity also extends into regions that are currently solely constrained by astrophysical observations (e.g., HB stars). 

We make some final comments on the complementarity of our work with respect to constraints from astrophysics. Generally, such constraints depend on certain underlying astrophysical assumptions: for supernovas, these assumptions involve details of the core-collapse simulation, while for stars they depend on standard stellar models and evolutionary stellar timescales. While it is unlikely that these astrophysical assumptions would be relaxed to the level where PASSAT would become competitive with astrophysical constraints, we point out that there has been recent work that hints at this possibility, at least for supernovae (for example \cite{Bar:2019ifz}). A detailed investigation of the astrophysics is beyond the scope of this paper. Our focus in this paper, instead, has been to entertain the following possibility: \textit{even} if one follows standard astrophysics and takes results from both SN1987A and HB stars, then it is \textit{still possible} to construct ALP models such that stellar or supernova environments are blind to the ALP-photon coupling. Thus, the relaxation of bounds is not due to astrophysics but rather due to hidden sector model building. Models which evade all stellar bounds (including energy loss arguments) have in fact been constructed in the literature and were actively studied in the wake of the PVLAS anomaly several years ago. Generally, such models proceed by introducing hidden sector dynamics that switches off the Primakoff production in stellar environments, either through appropriate choices of hidden sector charges and couplings \cite{Jaeckel:2006xm}, or through the introduction of phase transitions \cite{Mohapatra:2006pv}, screening mechanisms \cite{Brax:2007ak}, or form factors \cite{Ahlers:2006iz}.  We do not want to judge whether these models are natural; however, they do motivate the complementary laboratory-based approach based on PASSAT that we have pursued.

The models constructed to evade astrophysical bounds typically make the ALP production mechanism depend on the ambient plasma mass and high temperature environment in the interior of the star. Thus, the properties of the plasma become important, and the calculation must be performed after taking into account thermal  corrections. In the case of PASSAT, the production mechanism is simply associated with the usual beam dump environment where there are no effects from an ambient plasma. Thus, models created to evade astrophysical bounds and account for the PVLAS anomaly by invoking plasma/thermal physics would continue to account for a possible anomaly at PASSAT and evade CAST. We note, however, that the Fermi energy and electron density of the target material in which ALPs are produced in PASSAT may have important effects on such models, compared to typical light shining through wall experiments where ALPs are produced in vacuum.

In models where the form factor of the ALP-photon vertex is assumed to be momentum-dependent, the strategy to evade astrophysical bounds and account for the PVLAS anomaly models the form factor to attenuate for high momentum ($\sim$ keV inside stars) and strengthen at low momentum (for lasers). In our case, such form factor-dependent models would have to exhibit a different behavior: they would have to strengthen at high momentum ($\sim$ GeV that is relevant for PASSAT) and attenuate at low momentum ($\sim$ keV and below, to evade astrophysical and LSW bounds). Whether or not models with such form factors are natural would depend on the details of the model-building. Furthermore, since ALPs are produced in a target material in PASSAT, the Fermi energy and non-negligible electron density may become important when considering such models.

Finally, we note that our  beam production and background are in estimation level; a full simulation is being performed for an upcoming paper~\cite{future}. We believe that our estimations are conservative and a full treatment will lead to more robust constraints on the ALP parameter space. We also emphasize that the idea of PASSAT can be  implemented at other facilities and beam lines (e.g., electron beam dumps).
We reserve these topics for future work.

\section*{Acknowledgments} 
We would like to thank Babette D\"{o}brich, Kwang-Sik Jeong, and Pierre Sikivie for their careful reading of the draft.
We would also like to thank Bhaskar Dutta, Ahmed Ismail, Lucien Heurtier, Felix Kling, Jong-Chul Park, Seodong Shin, and Scott Thomas for useful comments and discussions.
D.K. acknowledges the hospitality of the Aspen Center for Physics, which is supported by National Science Foundation grant PHY-1607611, while completing this work. 
The work of D.K. was supported in part by the Department of Energy under Grant DE-FG02-13ER41976 (de-sc0009913) and is supported in part by the Department of Energy under Grant de-sc0010813. 
The work of K.S. is supported by the Department of Energy under Grant DE-SC0009956.

\bibliographystyle{apsrev4-1}
\bibliography{biblio}

\end{document}